\pdfoutput=1 
\RequirePackage[mathlines]{lineno}
\documentclass[onecolumn,10pt, notitlepage,aps,prb,superscriptaddress,floatfix,preprintnumbers,amsmath,amssymb]{revtex4}

\usepackage[a4paper, total={6in, 8in}]{geometry}
\usepackage{graphicx}

\usepackage{dcolumn}
\usepackage{bm}
\usepackage{subfigure}

\usepackage{xr}
\makeatletter
\newcommand*{\addFileDependency}[1]{
  \typeout{(#1)}
  \@addtofilelist{#1}
  \IfFileExists{#1}{}{\typeout{No file #1.}}
}
\makeatother

\begin{document}

\title{Domain wall patterning and giant response functions in ferrimagnetic spinels}

\author{L. L. Kish}
\email{lazark2@illinois.edu}
\affiliation{Department of Physics and Materials Research Laboratory, University of Illinois at Urbana-Champaign, Urbana, Illinois 61801, USA}

\author{A. Thaler}
\affiliation{Department of Physics and Materials Research Laboratory, University of Illinois at Urbana-Champaign, Urbana, Illinois 61801, USA}
\affiliation{Neutron Scattering Division, Oak Ridge National Laboratory, Oak Ridge, Tennessee 37831, USA}

\author{M. Lee}
\affiliation{National High Magnetic Field Laboratory, Los Alamos National Laboratory, Los Alamos, New Mexico 87544, USA}

\author{A. V. Zakrzewski}
\affiliation{Department of Physics and Materials Research Laboratory, University of Illinois at Urbana-Champaign, Urbana, Illinois 61801, USA}

\author{D. Reig-i-Plessis}
\affiliation{Department of Physics and Astronomy and Quantum Matter Institute, University of British Columbia, Vancouver, BC V6T 1Z1, Canada}

\author{B. Wolin}
\affiliation{Department of Physics and Materials Research Laboratory, University of Illinois at Urbana-Champaign, Urbana, Illinois 61801, USA}

\author{X. Wang}
\affiliation{Department of Physics and Materials Research Laboratory, University of Illinois at Urbana-Champaign, Urbana, Illinois 61801, USA}

\author{K.C. Littrell}
\affiliation{Neutron Scattering Division, Oak Ridge National Laboratory, Oak Ridge, Tennessee 37831, USA}

\author{R. Budakian}
\affiliation{Department of Physics and Materials Research Laboratory, University of Illinois at Urbana-Champaign, Urbana, Illinois 61801, USA}
\affiliation{Department of Physics and Astronomy, University of Waterloo, Waterloo, Ontario N2L 3G1, Canada}

\author{H. D. Zhou}
\affiliation{Department of Physics and Astronomy, University of Tennessee, Knoxville, Tennessee 37996, USA}

\author{V. S. Zapf}
\affiliation{National High Magnetic Field Laboratory, Los Alamos National Laboratory, Los Alamos, New Mexico 87544, USA}

\author{A. A. Aczel}
\affiliation{Neutron Scattering Division, Oak Ridge National Laboratory, Oak Ridge, Tennessee 37831, USA}

\author{L. DeBeer-Schmitt}
\affiliation{Neutron Scattering Division, Oak Ridge National Laboratory, Oak Ridge, Tennessee 37831, USA}

\author{G. J. MacDougall}

\affiliation{Department of Physics and Materials Research Laboratory, University of Illinois at Urbana-Champaign, Urbana, Illinois 61801, USA}

\date{\today}

\begin{abstract}

The manipulation of mesoscale domain wall phenomena has emerged as a powerful strategy for designing ferroelectric responses in functional devices, but its full potential has not yet been realized in the field of magnetism. We show that mechanically strained samples of Mn$_3$O$_4$ and MnV$_2$O$_4$ exhibit a stripe-like patterning of the bulk magnetization below known magnetostructural transitions, similar to the structural domains reported in ferroelectric materials. Building off our previous magnetic force microscopy data, we use small angle neutron scattering to  show that these patterns extend to the bulk, and demonstrate an ability to manipulate the domain walls via applied magnetic field and mechanical stress. We then connect these domains back to the anomalously large magnetoelastic and magnetodielectric response functions reported in these materials, directly correlating local and macroscopic measurements on the same crystals.

\end{abstract}

\maketitle


\section{Introduction}

Domain walls~\cite{Hubert2009,Salje2012,Wu2015}, dislocations~\cite{Mitchell2004}, and other moving defects~\cite{KakeshitaTomoyukiFukudaTakashiSaxenaAvadh2012} are critically important to the response functions of applied materials. One notable example is the nanometer-lengthscale domain wall patterning regularly observed in ferroelastic and ferroelectric materials as a means to reduce the energy associated with long-range strain fields~\cite{Salje2012,Feigl2014,Streiffer2002}. In ferroelectrics, these domain walls can be charged and conducting~\cite{Bednyakov2018}, and can drastically alter the macroscopic conductive and ferroelectric response functions~\cite{Liu2017}. As such, the control of domain patterning in these materials has been extensively examined in pursuit of new optoelectronic and capacitive memory devices~\cite{Li2008}.

Coexisting magnetic degrees of freedom add exciting new possibilities~\cite{Spaldin2019}. Single phase~\cite{Cheong2007} and composite~\cite{Nan2008} multiferroics have concomitant electric and magnetic moments, eliciting great interest from the spintronics community. Much like pure ferroelectrics, multiferroic materials have demonstrated mesoscale domain patterning~\cite{Wu2015}, with conductive domain walls~\cite{Rojac2017,Bencan2020} and magnetoelectric couplings~\cite{Cheong2007} opening possibilities for field-tunable conductive or capacitive devices~\cite{Spaldin2019,Nan2008}. An enhanced magnetodielectric response has often been associated with magnetic transitions~\cite{Lawes2009}, and is one signature of coupled ferroelectric and ferromagnetic orders~\cite{Cheong2007}. However, even in absence of ferroelectricity, enhanced magnetodielectric couplings can arise through local spin-phonon coupling or extrinsic mechanisms\cite{Lawes2009,Catalan2006}.

Prime examples are found in the literature on ferrimagnetic spinel oxides. These materials are notable for strong spin-lattice coupling and the prevalence of low-temperature magnetostructural transitions~\cite{Lee2010} into states often characterized by giant magnetoelastic and magnetodielectric response functions~\cite{Mufti2010,Liu2020,Tackett2007,Kemei2014,Suzuki2007,Suzuki2008}. Recent studies by some of us using magnetic force microscopy (MFM)~\cite{Wolin2018} and others using transmission electron microscopy (TEM)~\cite{Murakami2011,Kasama2010} have revealed the existence of stripe-like magnetization domains in the same temperature region.

In this article, we present a small-angle neutron scattering (SANS) study exploring the distribution and field response of these magnetization domain patterns in two specific spinel materials: $\mathrm{Mn_3O_4}$ (MMO) and $\mathrm{MnV_2O_4}$ (MVO). MMO is the prototypical tetragonal spinel, with a Jahn-Teller transition at 1440~K~($\mathrm{T^{MMO}_{JT}}$)~\cite{DORRIS1988}, and three magnetic transitions~\cite{Jensen1974, Chardon1986} at 42~K~($\mathrm{T^{MMO}_{FM}}$), 39~K~($\mathrm{T^{MMO}_{IC}}$), and 33~K~($\mathrm{T^{MMO}_{MS}}$). The 33~K transition was recently shown to have a structural component~\cite{Kim2011}.

MVO has a magnetic ordering transition at 56~K~($\mathrm{T^{MVO}_{FM}}$) and an orbital-ordering transition at 53~K~($\mathrm{T^{MVO}_{MS}}$), which increases coupling between spins and the lattice~\cite{Garlea2008,Suzuki2007}. At the lowest temperatures, MMO features a net ferrimagnetic moment along $\mathrm{b}$ and a macroscopic hard axis in the $\mathrm{c}$ direction~\cite{Kim2012}, while MVO in contrast has $\mathrm{c}$ as its easy axis.  (Supplementary Section~S-1 details our crystallographic conventions.)

In both materials, the lowest temperature phases have been associated with large magnetoelastic and magnetodielectric response functions~\cite{Suzuki2008,Tackett2007,Liu2020,Kismarahardja2013,Suzuki2008,Nii2012} and stripe-like magnetic domains in our previous study~\cite{Wolin2018}. Our current SANS data build upon our initial MFM results and reveal the extension of low-temperature stripe-like domains to the bulk of each material. We further show a strong response of these walls to applied magnetic fields, tunable with applied stress and strongly correlated with magnetization and magnetodielectric response functions.

\section{Characterization of domain structure}\label{section:domains}
\begin{figure*}[!htpb]
	\centering
	\includegraphics[width=0.7\textwidth]{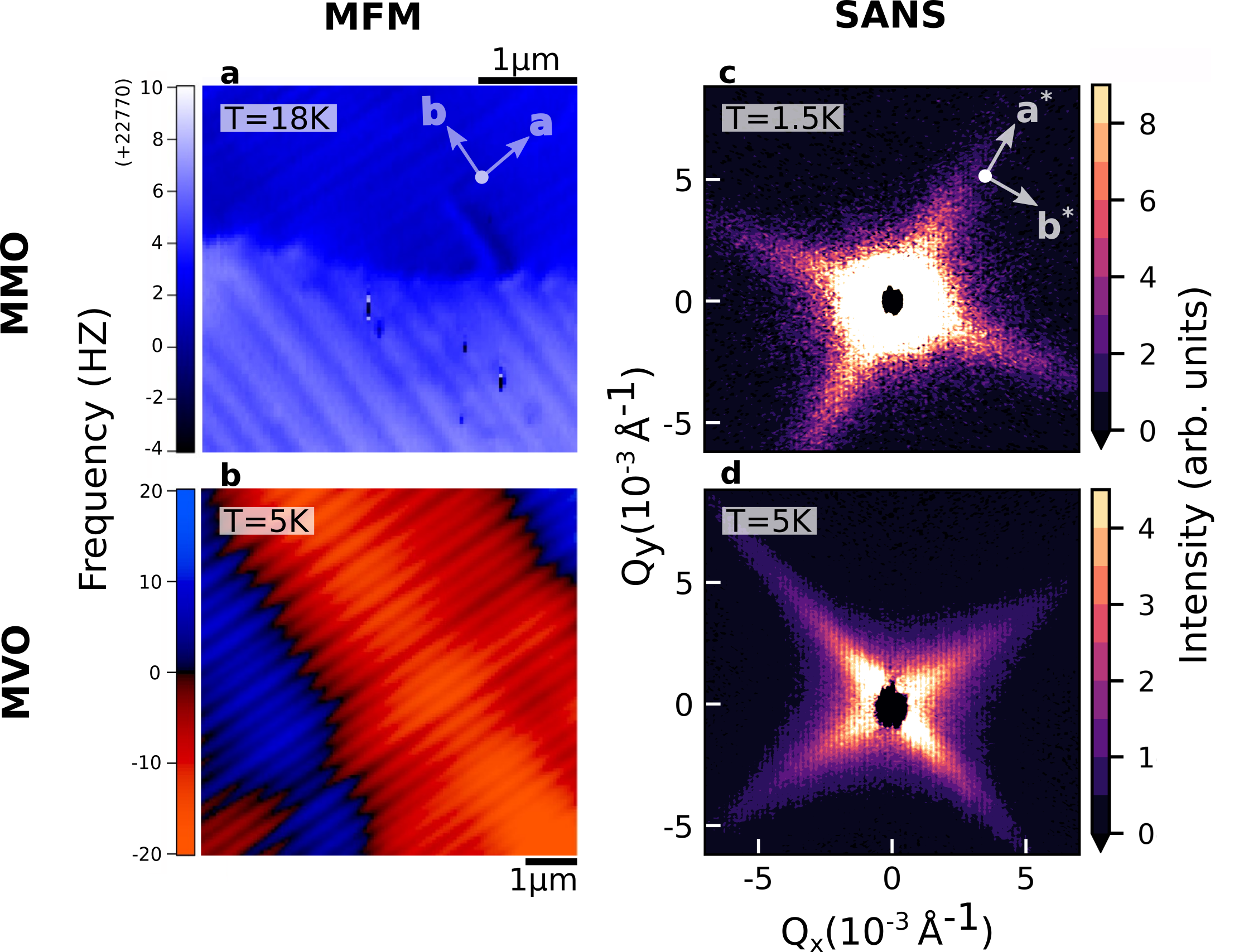}
	\caption{\textbf{Magnetic domain structure via MFM and SANS.} \textbf{a},\textbf{b} MFM images of stripe patterning on the (00l) surfaces of (\textbf{a}) MMO and (\textbf{b}) MVO, taken below respective structural transitions. Reported MVO frequency is shift with respect to natural resonance frequency of the probe. \textbf{c},\textbf{d} SANS patterns from the tetragonal (hk0) scattering plane of (\textbf{c}) MMO at $\mathrm{T=1.5K}$ and (\textbf{d}) MVO at $\mathrm{T=5K}$, with the measured scattering in the paramagnetic phase subtracted off as background.}
	\label{fig:stripesFins}
\end{figure*}

In Fig.~\ref{fig:stripesFins}~a~(b), we show real-space MFM images of zero field-cooled magnetic domain patterns below the respective magnetostructural transitions in MMO~(MVO) single crystals, previously published by us in Ref.~[\onlinecite{Wolin2018}]. Fig.~\ref{fig:stripesFins}~c~(d) show directly comparable SANS intensities in MMO~(MVO) at 1.5~K~(5~K) and zero field, after subtraction of high-temperature paramagnetic patterns. On the surface of each material, MFM reveals 100 nm lengthscale modulations of the underlying ferrimagnetism propagating parallel to the tetragonal $\mathrm{a/b}$ axes. The morphology and oriented nature of these domains imply a coupling to the crystal structure akin to ferroelastic domains~\cite{Salje2012}. 

Our neutron scattering data show that the stripe domains are not surface effects, but persist through the bulk of our single-crystal samples. The major features can be mapped directly to the hierarchical magnetic domain structure observed in the real-space MFM maps. The stripe patterns manifest distinctly as anisotropic fins of intensity stretching along $\mathrm{a^*}$ and $\mathrm{b^*}$ directions, the same directionality as the MFM domain walls and, as shown below, reflecting the same characteristic lengthscales.

In MMO, the scattering also contains an isotropic component, which we associate with the 1-10 micron domains that typically form in ferromagnetic materials to minimize demagnetization energy\cite{Hubert2009, Muhlbauer2019}. Signatures of this are less prevalent in SANS patterns for MVO, partially due to the larger size of the dipolar domains as shown by MFM\cite{Wolin2018}.

\begin{figure*}
	\centering
	\includegraphics[width=.7\textwidth]{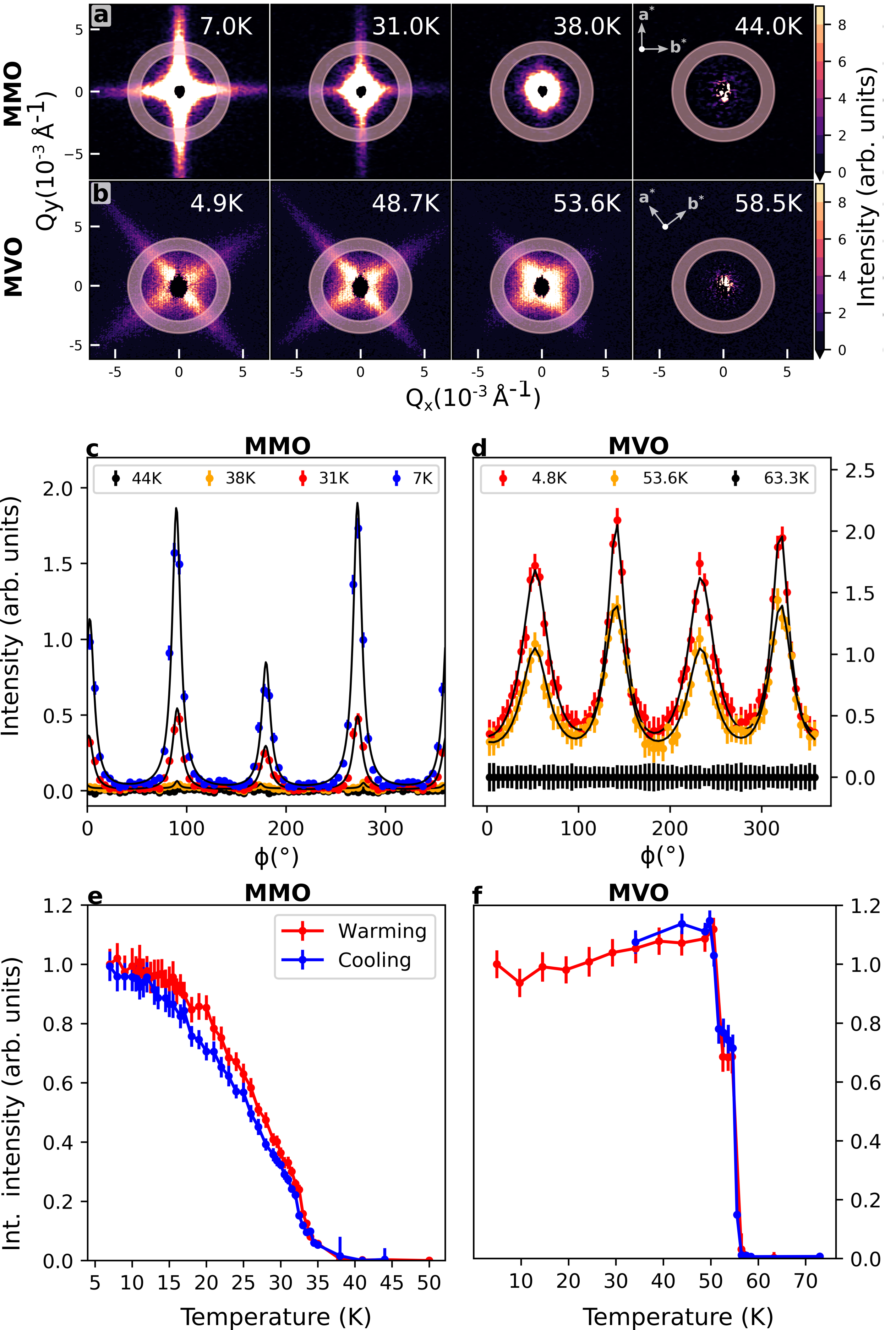}
	\caption{\textbf{Temperature dependence of SANS intensity.} \textbf{a},\textbf{b},~Raw scattering intensity from SANS in the tetragonal (hk0) plane at selected temperatures in (\textbf{a})~MMO and (\textbf{b}) MVO. \textbf{c},\textbf{d},~Annular Q-cuts in (\textbf{c})~MMO and (\textbf{d})~MVO, with an integration range of 0.003-0.004~\AA$^{-1}$ with the associated Lorentzian fits superimposed on the data. \textbf{e},\textbf{f},~Integrated intensity of the fin scattering plotted for (\textbf{e})~MMO and (\textbf{f})~MVO as a function of temperature on both warming and cooling. Error bars represent one standard deviation in panels \textbf{c,d} and one standard error as determined from nonlinear least squares fitting in panels \textbf{e,f}.}
	\label{fig:IvT}
\end{figure*}

Fig.~\ref{fig:IvT}~a-b show the progression of the 2D SANS patterns in each material, as temperature is lowered below their respective ferrimagnetic ordering ($\mathrm{T_{FM}}$) and magnetostructural ($\mathrm{T_{MS}}$) transitions. Both show weak scattering near Q=0 at highest temperatures, consistent with paramagnetic scattering above an ordered state with a net magnetic moment. In MMO, isotropic scattering appears immediately below $\mathrm{T^{MMO}_{FM}}$, and the anisotropic features develop as a redistribution of this intensity below $\mathrm{T^{MMO}_{MS}}$. MVO features no temperature region with isotropic scattering. Instead, the anisotropic scattering appears immediately at $\mathrm{T^{MVO}_{FM}}$ and jumps in intensity at $\mathrm{T^{MVO}_{MS}}$, with the intermediate region between these two transitions characterized by a shift in scattering toward lower $\mathrm{Q}$. MFM reports that this region is characterized by sporadic stripe development at zero field~\cite{Wolin2018}, consistent with a cubic-tetragonal phase mixture, as has been reported elsewhere~\cite{Zhou2007}.

To quantify these trends, annular cuts were taken for $\mathrm{Q}$ from 0.003-0.004~\AA$^{-1}$ (the shaded regions in Fig.~\ref{fig:IvT} a-b) and fit to Lorentzian transverse profiles, as shown in Fig.~\ref{fig:IvT} c-d. The integrated intensities of the Lorentzian fits are plotted in the lower panels on both warming and cooling. The results show that the fins appear in MMO at $\mathrm{T^{MMO}_{FM}}$ and saturate below 10K, and appear in MVO in two sharp steps at $\mathrm{T^{MVO}_{FM}}$ and $\mathrm{T^{MVO}_{MS}}$. These trends and the weak temperature hysteresis are largely in line with a collection of Raman spectroscopy, diffraction, and macrostrain measurements in the literature tracking the magnetostructural transitions in each material~\cite{Suzuki2007,Nii2012,Garlea2008,Suzuki2008,Byrum2016}.

\begin{figure*}
	\centering
	\includegraphics[width=.7\textwidth]{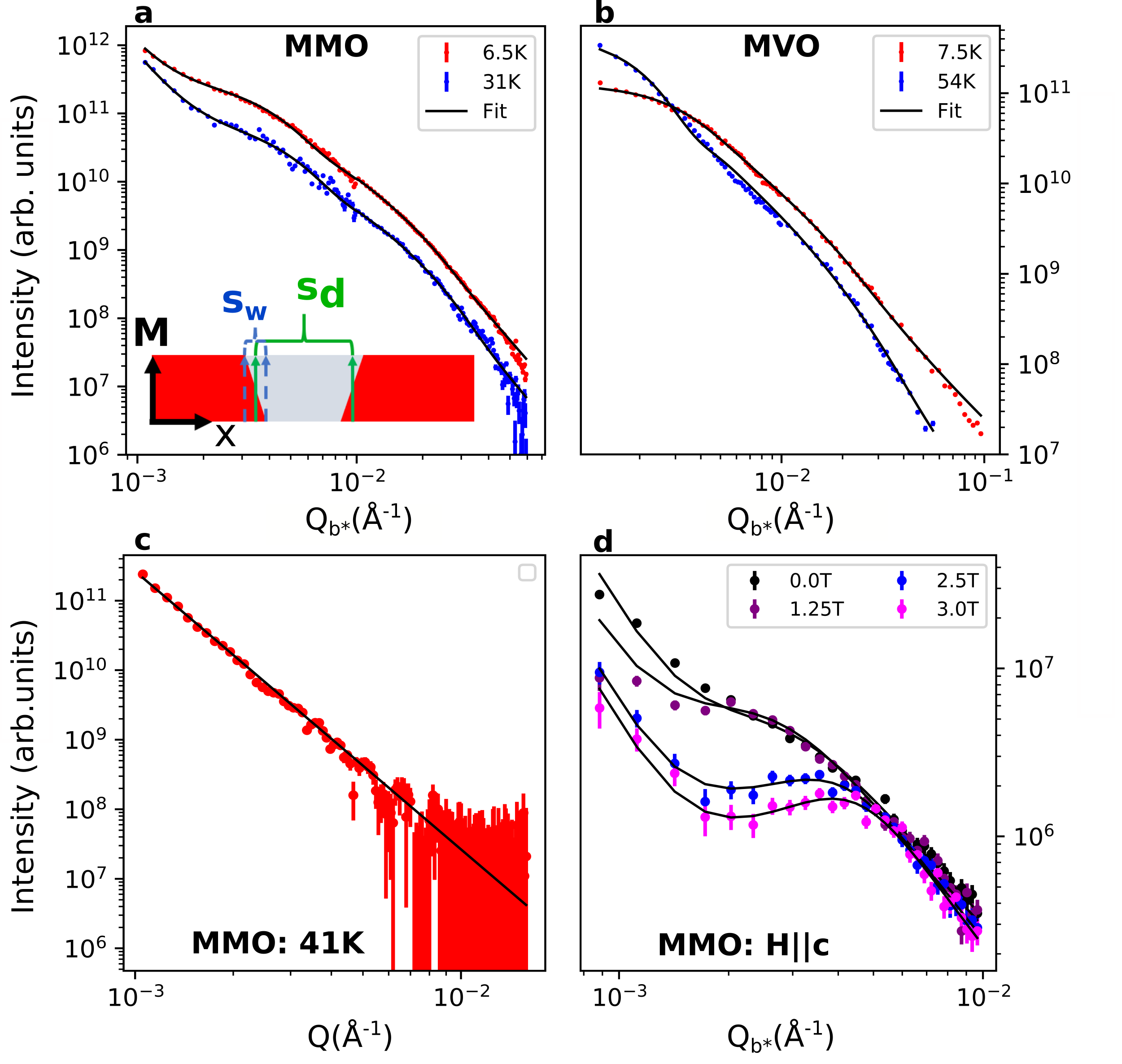}
	\caption{\textbf{SANS intensity momentum dependence.} \textbf{a},\textbf{b} Dependence of SANS intensity on $\mathrm{Q_{b^*}}$ for (\textbf{a})~MMO and (\textbf{b})~MVO with fits to the 1D stripe model described in the text superimposed on the data. The inset of \textbf{a} shows a real-space schematic of the model, with stripe domain size and wall size $\mathrm{s_d}$ and $\mathrm{s_w}$. \textbf{c} Isotropic average of intensity versus $\mathrm{Q}$ in MMO at 41~K, showing Porod scattering arising from dipolar domains. \textbf{d} Field-induced redistribution of MMO low-$\mathrm{Q}$ intensity versus $\mathrm{Q_{b^*}}$. Error bars represent one standard deviation.}
	\label{fig:IvQ}
\end{figure*}

Fig.~\ref{fig:IvQ} shows the dependence of the scattering intensity on momentum $\mathrm{Q}$ in various directions for both materials. Figs.~\ref{fig:IvQ}~a-b show intensity profiles of rectangular line cuts along $\mathrm{b^*}$ at temperatures where fin scattering is visible, with the transverse width varying from 0.002-0.016 \AA$^{-1}$ to capture the full breadth of the features using the different instrument settings (Supplementary Fig.~S-2. These profiles are notable for their subtle curvature on a log-log scale, which can be captured by a simple model (solid line) described below. For MMO, we also observe characteristic Porod scattering ($\mathrm{I=b^*Q^{-4}}$)~\cite{Muhlbauer2019} in the low-$\mathrm{Q}$ region of our data, which is typical of ferromagnetic dipolar domains~\cite{Muhlbauer2019} with smooth and isotropically distributed walls. This power law behaviour accounts for all of the intensity in the ferrimagnetic temperature region between $\mathrm{T^{MMO}_{FM}}$ and $\mathrm{T^{MMO}_{MS}}$, as demonstrated by the isotropically averaged 41~K dataset displayed in Fig.~\ref{fig:IvQ}~c.

The $\mathrm{Q}$-dependence of the scattering intensity along the fins was captured by the simple model equation:
\begin{linenomath*}
    \begin{equation}
    \mathrm{I(Q_{b^*})=a\left(sinc^2\left(\frac{s Q_{b^*}}{2}\right) * G(s,s_d,\sigma_d)\right)\left(sinc^2\left(\frac{s Q_{b^*}}{2}\right) * G(s,s_w,\sigma_w)\right).}
    \end{equation}
\end{linenomath*}
This model supposes that individual stripe domains are one-dimensional regions of constant magnetization, with linear domain walls. The domains and their walls have variable widths of $\mathrm{s_d}$ and $\mathrm{s_w}$, respectively, and $\mathrm a$ is a scaling parameter. The 1D domain form factors are convoluted with normalized Gaussian size distributions
\begin{linenomath*}
    \begin{equation}
    \mathrm{G=\frac{\exp(-\frac{(s-s_*)^2}{2\sigma^2})}{\sqrt{2\pi}\sigma}}
    \end{equation}
\end{linenomath*}
with center $\mathrm{s_*}$~$=$~$\mathrm{s_d}$ ($\mathrm{s_w}$) and width $\mathrm \sigma$~$=$~$\mathrm{\sigma_d}$ ($\mathrm{\sigma_w}$) to account for polydispersity. Additional details are listed in Supplementary Section~S-4. 

Fits to this model at base temperature provide a mean stripe width of 86.7 $\mathrm\pm$ 1.7 nm for MMO and 90.5 $\mathrm\pm$1.3 nm for MVO, which are consistent with the local MFM observations~\cite{Wolin2018}. The domain walls were found to have a width 15.8 $\mathrm\pm$ 0.2 nm for MMO and 15.8 $\mathrm\pm$ 0.5 nm for MVO, below the resolution limit of MFM. In the intra-transition region of MVO~(T=54~K), the intensity distribution of the fin scattering moves closer to Q=~0 and fits correspondingly yield a larger mean stripe width of 169.0 $\mathrm\pm$ 3.0 nm with a 13.3 $\mathrm\pm$ 1.4 nm wall width. Standard deviations of the size distributions converged to quite large values ($\mathrm{\sigma_d\approx30\%}$ of the mean size), explaining in part the lack of strong intensity peaks in our data. Though the intensity distribution in MMO was well fit, our estimate of the domain wall width should be thought of as an upper bound due to complications with grain misalignments in this material. 

\begin{figure*}[!htpb]
	\centering
	\includegraphics[width=0.7\textwidth]{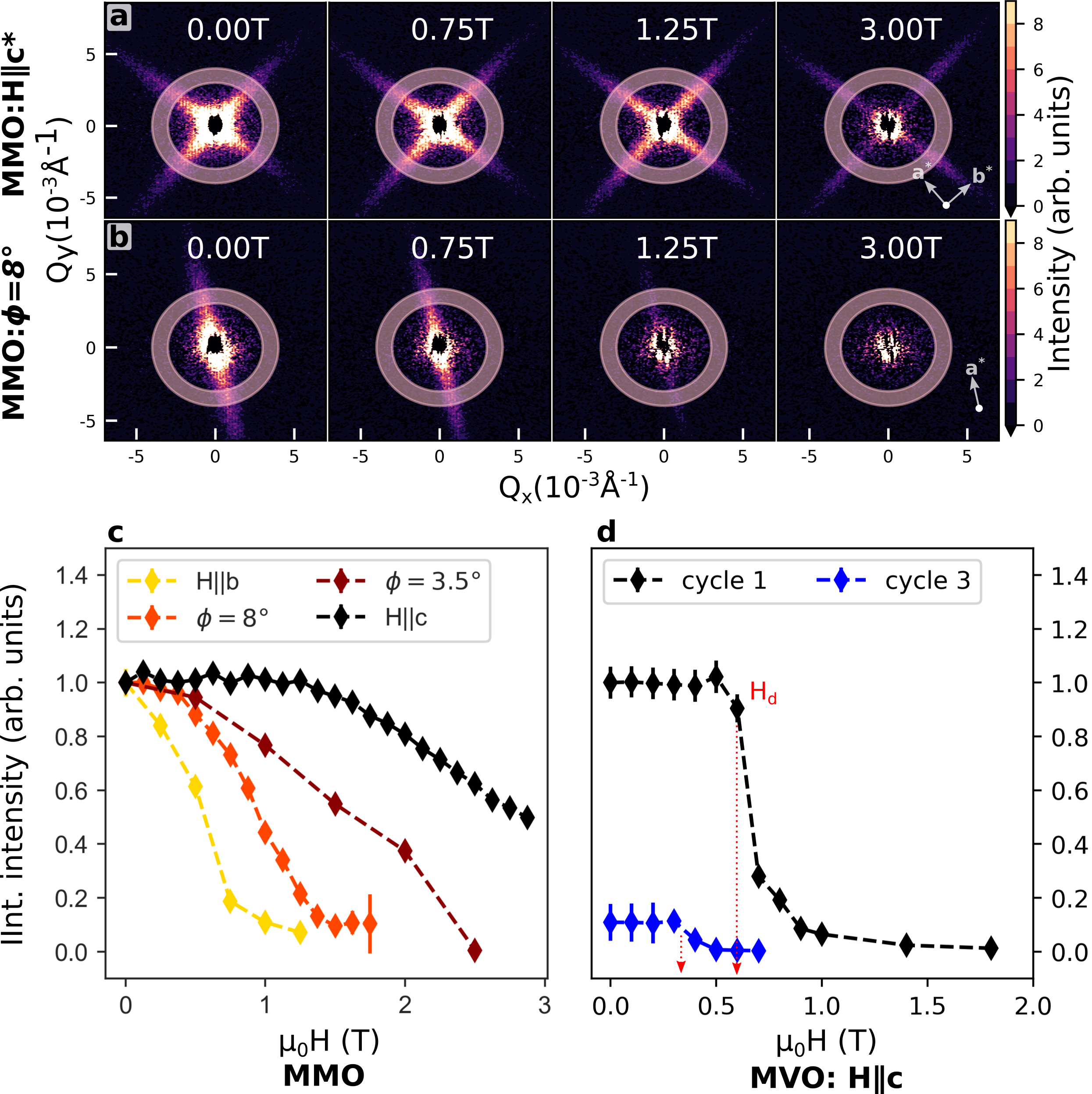}
	\caption{\textbf{Domain response to applied magnetic field in SANS.} \textbf{a},\textbf{b}, Raw 2D SANS data from MMO as a function of increasing field (\textbf{a}) parallel to the hard $\mathrm{c}$ axis and (\textbf{b})~$\mathrm{\phi=8^{\circ}}$ misaligned from $\mathrm{c}$, demonstrating anisotropic response in the fin intensity. \textbf{c} MMO integrated fin intensity in the range 0.003-0.004~\AA$^{-1}$ as a function of field applied along different directions with respect to $\mathrm{c}$. Intensities were rescaled to match the first datapoint between orientations. \textbf{d} MVO integrated fin intensity in the range 0.01-0.12~\AA$^{-1}$ as a function of field along $\mathrm{c}$ axis normal to sample plate, for two field cycles as described in the text, with equal scaling for the intensities. Error bars in panels \textbf{c,d} represent one standard error as determined from nonlinear least squares fitting.}
	\label{fig:IvH}	
\end{figure*}

\section{Response to applied field}\label{section:field}

A notable feature of the stripe-like domains is their responsivity to applied magnetic field. This can be seen in Fig.~\ref{fig:IvQ}~d, which shows the evolution of $\mathrm I$ vs $\mathrm{Q_{b^*}}$ in MMO at base temperature as magnetic fields are applied in the out-of-plane $\mathrm{c}$-axis direction. In contrast to relatively featureless low field curves, applied fields of a few Tesla redistribute the scattering intensity until peaks representing distinct spatial correlations are visible. This corresponds to the system evolving from a state with weak correlations to one where stripes are more regularly arranged. The position of the peak at $\mathrm{H}$=3T corresponds to inter-stripe spacings of approximately 300 nm, comparable to distances observed with MFM~\cite{Wolin2018}.

Representative 2D scattering patterns for MMO at $\mathrm{T =}$~2K are shown in Fig.~\ref{fig:IvH}~a-b for measurements performed both with $\mathrm{H \parallel c}$ and with the sample tilted an angle $\mathrm{\phi=8^\circ}$ off-axis. For the former, the most obvious effect of applied field is the above-described sharpening of the fin features, with only a moderate decrease in integrated intensity at the highest field available (3T). However, if field direction is varied even a few degrees, fin intensity decreases much more rapidly and disappears almost entirely by 1T. This strong anisotropy presents more clearly in Fig.~\ref{fig:IvH}~c, where we show integrated scattering intensity in the fins versus $\mathrm H$ for four different field orientations. Subsequent removal of applied fields shows no return of fin scattering. These observations suggest an anisotropic depinning field $\mathrm{H_d}$, above which stripe domain walls respond to increase the sample magnetic moment. Similar depinning fields have been reported for structural domains in ferroelectrics~\cite{Rojac2010,Bencan2020}. This interpretation agrees with MFM studies on MMO~\cite{Wolin2018,WangDissertation} and our own magnetization data below.

Fig.~\ref{fig:IvH}~d demonstrates a similar field effect in MVO, with field along the easy $\mathrm{c}$ axis normal to the mounting plate. A well-defined initial plateau yields to a steep drop of integrated fin intensity at H=$\mathrm{H_d}$ with increasing field. In an interesting contrast to MMO, field and temperature cycling of the sample appears to lower both $\mathrm{H_d}$ and the integrated fin intensity in MVO, as can be seen in the difference between cycles 1 and 3 (spacing in field for cycle~2 was insufficient for a useful comparison). In one ``cycle" here, the sample is zero-field cooled to base temperature, whereupon field is ramped to a high value ($\mathrm{>1.2T}$) during measurement. Field is subsequently removed, and the sample heated above its magnetic ordering temperature ($\mathrm{>60K}$) before cooling for further scans. Despite thermally cycling above the transition, the low temperature magnetic properties still retain memory about the number of cycles; this can be understood in the context of sample strain, discussed further below. 

\begin{figure*}[htp]
	\centering
	\includegraphics[width=0.7\textwidth]{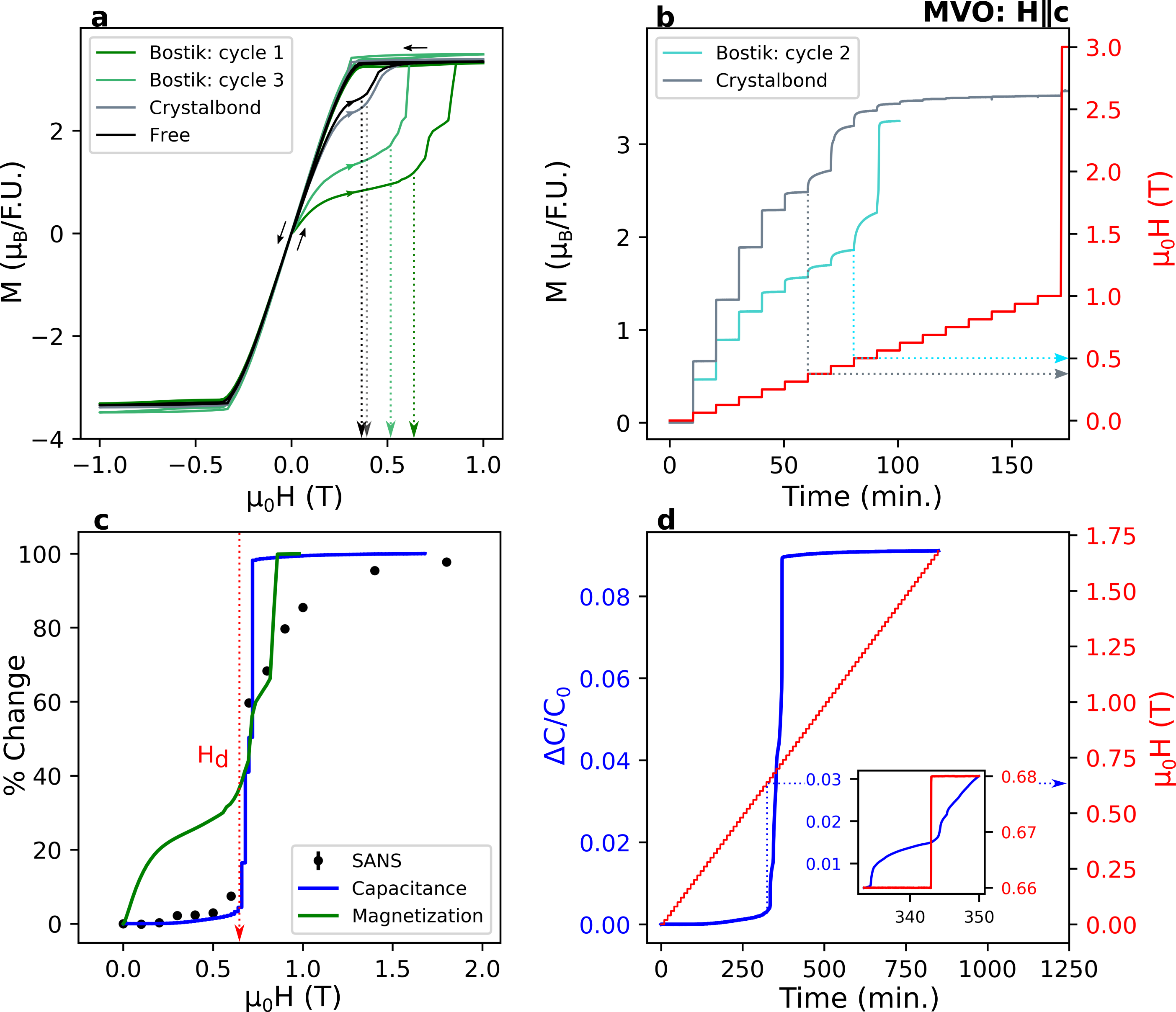}
	\caption{\textbf{Connection between stripe domains and bulk response functions.} Dashed lines indicate the inferred location of $\mathrm{H_d}$. \textbf{a} Magnetic hysteresis behavior of MVO at 2~K under various mounting conditions with sample either allowed to freely contract, or glued to mount with Crystalbond adhesive or Bostik superglue. Two Bostik field-temperature cycles as described in the main text are shown. \textbf{b} Time-dependent magnetization in which field is ramped in 60~mT steps to observe slow relaxations associated with domain wall motion. \textbf{c} Magnetic field-induced change in the quantities measured by our different probes (i.e. SANS integrated fin intensity, virgin magnetization, capacitance), plotted as a percentage of total change versus field. \textbf{d} Measurement of sample capacitance undergoing time dependent field-ramps (10~mT steps), plotted as a fractional difference ($\mathrm{\frac{\Delta C}{C_0}}$) from the zero field-cooled capacitance value ($\mathrm{C_0}$). The inset shows the slow relaxations we associate with low-temperature stripe domain wall motion.}
	\label{fig:Bulk}
\end{figure*}

\section{Comparison to bulk response functions} \label{section:bulk}

To understand the effect of in-field stripe behaviour on macroscopic response functions, we correlate our SANS data with separate measurements of both magnetization and magnetocapacitance on the same crystalline samples. A subset of this data for MVO is shown in Fig.~\ref{fig:Bulk}, and as discussed below, connects the presence and motion of mesoscale stripe domain walls with anomalous magnetic responses in both spin and charge channels.

In Figs.~\ref{fig:Bulk}~a-b, we show the magnetization of MVO at 2~K with field $\mathrm{\mathbf{H}\parallel c}$ normal to the surface of the sample mount. Fig.~\ref{fig:Bulk}~a displays applied field hysteresis curves for mounting configurations with the crystal glued to the sample mount with various commercially available adhesives or held in place by teflon tape and free to expand or contract. In all configurations, zero field-cooled initial (``virgin") magnetization curves show an anomaly at finite field, above which there is a more rapid positive change in response. When attached with Bostik superglue (green), the anomaly occurs at $\mathrm{\mu_0 H_d\sim 0.6 T}$, which is the same field where fin scattering begins to decrease in SANS measurements under similar mounting conditions. When mounted with the adhesive Crystalbond (gray), both $\mathrm{H_d}$ and the size of the virgin hysteresis loop are suppressed. The freely mounted sample (black) shows marginal reduction of the hysteresis loop as compared with Crystalbond. Subsequent cycles of temperature and field systematically reduce the Bostik virgin hysteresis loop, as shown in Fig.~\ref{fig:Bulk}~a.

In all mounting configurations, data reveal a crossover in the timescale of the magnetization response near $\mathrm{H=H_d}$. This is seen in  Fig.~\ref{fig:Bulk}~b, which shows magnetization versus time (gray/cyan) as applied field is ramped up in a series of sharp steps (red). Below $\mathrm{H_d}$, the magnetization approaches steady state at a rate faster than the time resolution of the instrument ($<$1s). Above $\mathrm{H_d}$, where stripe domain wall motion is inferred from SANS, the response proceeds with a slower characteristic timescale, typically tens of minutes. Once saturated from the virgin state, further hysteresis loops are typical of a soft ferromagnet, with little coercivity, remnant magnetization, or appearance of slow relaxation timescales. Comparable data for MMO can be found in the Supplementary Section~S-6.

In Fig. ~\ref{fig:Bulk}~c, we overplot the change in integrated fin intensity from SANS with Bostik cycle~1 virgin magnetization as a percentage of the total change. If one associates the virgin deviation in magnetization with the pinning of domain walls, then this plot indicates that bulk magnetic properties of our materials are determined primarily by the motion of quasi-1D stripe domains, rather than micron sized dipolar domains as is typical for classical ferromagnets~\cite{Hubert2009}. 

Perhaps more significantly, similar correlations are found between stripe domains and the dielectric response. The blue curve in Fig.~\ref{fig:Bulk}~c represents the magnetocapacitance on the same MVO sample at T=3~K, measured using the conventional parallel-plate capacitor method with electric and magnetic fields aligned along $\mathrm{c}$. 
Much like magnetization, magnetocapacitance exhibits a distinct jump at H=$\mathrm{H_d}$. Fig.~\ref{fig:Bulk}~d displays time-dependent measurements revealing slow relaxations with a timescale comparable to magnetization for $\mathrm{H} > \mathrm{H_d}$. Moreover the field dependence of the virgin curve exhibits a $\mathrm{9.1\%}$ relative increase in capacitance, 60 times larger than subsequent hysteresis loops where SANS and magnetization signatures of the stripes are also removed. The hysteresis loops nevertheless show some dependence on the ramp-rate of the field, implying remnant time-dependent domain behavior.  Complementary capacitance data showing in-phase and out-of-phase capacitance signals over the entire hysteresis loop, as well as confirmation of a large sample resistance ($\>$6 T$\mathrm\Omega$ below 20K) are reported in Supplementary Section~S-8.

These measurements directly connect magnetic stripe domains with magnetization and magnetocapacitive effects, but cannot unambiguously disentangle capacitive contributions from magnetoelastic sample contraction and purely magnetodielectric effects. Published macrostrain and diffraction measurements~\cite{Garlea2008,Suzuki2007,Nii2012} allow us to place an upper bound on changes in the sample dimensions, often attributed to the redistribution of tetragonal domains. In the extreme case of a single domain crystal (lattice constants a=~8.53~\AA~and c=~8.42 \AA)~\cite{Suzuki2007} with $\mathrm{c}$ perpendicular to field switching entirely to a different single-domain state with $\mathrm{c}$ uniformly parallel with field, one would only expect a capacitance increase of $\mathrm{2.6\%}$. From this, we estimate an overall field-induced dielectric constant change of $\mathrm{>6.3\%}$ relative to the zero-field cooled value tied to the domain response.

The exact mechanism behind the observed magnetodielectric effects is unknown, but such phenomena can arise in  non-polar insulators from a coupling between optical phonons and local spins~\cite{Lawes2009}. At low temperature, MVO~\cite{Garlea2008,Gleason2014,Tabuko2011} develops orbital order on the vanadium sublattice, greatly enhancing the spin-lattice coupling in this material. One might expect that phonons coupling to the resulting antiferromagnetic modulation would produce a highly anisotropic dielectric tensor with respect to the $\mathrm{c}$-axis; field-induced rearrangement of the $\mathrm{c}$-axis domain structure would then yield large changes in the macroscopic capacitance far below the ordering transition.

Interfacial capacitances at the domain walls may also contribute, arising either from a spatially inhomogeneous dielectric tensor~\cite{Lunkenheimer2009,Catalan2006} or domain wall polarity due to bulk structural distortions~\cite{Yokota2017}. Our large measured sample resistance likely invalidates models invoking magnetoconduction~\cite{Catalan2006}, though this method cannot rule out an insulating surface layer. 

\section{The effect of strain}\label{section:strain}

Strain appears to be a key aspect of stripe domain formation in MVO and MMO. As reported in our previous publication~\cite{Wolin2018}, coordinated appearance of the stripe domains in maps of surface magnetization required the presence of mechanical stress as the samples proceed through their respective magnetostructural transitions, not inconsistently with bulk data in the literature~\cite{Suzuki2007,Nii2012,Tabuko2011}. This is also borne out by the magnetization data presented in Fig.~\ref{fig:Bulk}~a-b, where ``glued" virgin curves display suppressed initial susceptibilities and higher saturation fields. Both here and in Ref.~[\onlinecite{Wolin2018}], the strain in MVO crystals is presumed to be a result of differential thermal contraction between the sample and the sample mount, and the elastic properties and binding strength of the adhesive. Consequently, field and temperature cycling our sample without remounting systematically reduces the size of the magnetic hysteresis loop (Fig.~\ref{fig:Bulk}~a), as adhesive layers form microcracks to relieve strain fields. Similarly our SANS data (Fig.~\ref{fig:IvH}~d) show a notable decrease in both $\mathrm{H_d}$ and integrated intensities of fin scattering between the first and third field/temperature cycles for MVO.

The situation is more subtle in MMO, but likely closely related. For this material, neither SANS patterns nor bulk response functions are strongly dependent on mounting configuration, but vary strongly with the method employed to grow single crystals. Due to the presence of a Jahn-Teller transition at $\mathrm{T_{JT}}$=1440K, crystals grown at higher temperatures using the optical floating-zone method inherit an intrinsically strained twin structure at room temperature~\cite{Byrum2016}. MFM surface maps associate high strain regions in MMO with shorter-wavelength stripe modulations~\cite{Wolin2018}.

Low strain crystals obtained from flux growths at temperatures below $\mathrm{T_{JT}}$ demonstrate markedly different bulk properties~\cite{Byrum2016} and dramatically suppressed fin features in SANS patterns. Bulk magnetic and neutron characterization on the flux grown crystals are reported in Supplementary Sections~S-7 and S-8.

\section{Discussion and outlook}

This work shows that the 100~nm modulations in low-temperature ferrimagnetic phases drive the anomalous response behaviour of MMO and MVO . Atypical of ferromagnets, the lengthscale and directionality of these domains bear unmistakable similarity to the structural domains observed in high temperature ferroelectric materials. This suggests the stripes contain a structural component, with patterns forming to reduce long-range strain fields. The small elastic perturbations necessary to alter the domain structure point to stress as a useful control knob of response functions in these materials.

The unique characteristics of ferrimagnetic materials make them ideal candidates to demonstrate anomalous behaviour of this kind. Antiferromagnetic spin correlations couple strongly to optical phonons to enhance spin-lattice coupling~\cite{Lawes2009}, while the net ordered moment allows for coupling to external magnetic fields. 

Similar phenomenologies are expected for other materials with magnetostructural transitions, extending to other novel degrees-of-freedom and higher transition temperatures. One example is similar behavior observed below the Verwey transition ($\mathrm{T=125K}$) in magnetite~\cite{Kasama2010}, an inverse spinel multiferroic with unique local physics. Other ferrimagnetic spinels also display a broad selection of novel electronic properties~\cite{Yamasaki2006,MacDougall2012a} and pressure induced electron itinerancy~\cite{Kismarahardja2011,Reig-I-Plessis2016} which could couple to such domain effects. Our observations open up exciting new possibilities for tunable functionality in materials with spin-lattice coupling through manipulation of magnetostrucutal domains.

\section{Methods}

\renewcommand{\thepage}{M-\arabic{page}}
\renewcommand{\thesubsection}{M-\arabic{subsection}}
\renewcommand{\thetable}{M-\arabic{table}}
\renewcommand{\thefigure}{M-\arabic{figure}}

Single crystal samples of both materials were grown using the traveling-solvent floating zone method, as described in our previous work~\cite{Wolin2018, Gleason2014}. The samples were cut into a plate-like geometry with thickness ranging from 0.1 mm to 1 mm to reduce multiple scattering and absorption effects in SANS measurements. For MMO, additional crystals were grown at temperatures below the known cubic-tetragonal Jahn-Teller transition using anhydrous sodium tetraborate (Borax) flux, in order to reduce tetragonal domain formation and thus internal strain. SANS results on this low strain crystal show a substantial reduction in fin scattering, as shown in the Supplemental Material Fig.~S-7. Additional details regarding the two growth methods and the resulting strain effects for this material are given elsewhere\cite{Byrum2016}.

Our neutron scattering and magnetization measurements exhibit clear differences in the MVO domain structure depending on the strain induced by our choice of mounting method. Strained measurements were obtained by using commercially available adhesives Crystalbond and  Bostik superglue to attach samples to an aluminum backing plate. Low-strain magnetization measurements were obtained either by suspending the sample using teflon tape (held in place by the sample's plate-like geometry), enabling free thermal contraction.

MFM imaging was performed at the University of Illinois Materials Research Laboratory as detailed in our previous paper~\cite{Wolin2018}. Imaging of surfaces normal to $\mathrm{c}$ for each crystal at low temperature and under magnetic field are directly comparable to our SANS measurements. Samples were glued to copper mounts using Stycast 2850FT Epoxy, which created a finite sample strain comparable to those in our SANS measurements.

Our neutron scattering work was carried out at GP-SANS in the High Flux Isotope Reactor (HFIR) at Oak Ridge National Laboratory~\cite{Heller2018}. The bulk of our work was done with the neutron beam parallel to the tetragonal c-axis to probe the (hk0) plane of each material, and therefore the known stripe propagation direction. A series of discrete wavelengths between 4.76~$\text{\AA}$ and 19~$\text{\AA}$ with a $\frac{\Delta\lambda}{\lambda}$ of 0.13 was used to access $\mathrm{Q}$ between 0.001 $\text{\AA}^{-1}$ and 0.1 $\text{\AA}^{-1}$. As described in the Supplementary Section~S-3, finely-spaced rocking scans were obtained where necessary to account for the highly anisotropic nature of fin scattering in the out-of-plane direction of the detector~\cite{Bellet1992}. A horizontal field magnet was used to place magnetic field along the incident neutron beam, with temperature control achieved either by a variable temperature insert (base T=1.5~K) or a Helium-4 Closed-Cycle Refrigerator (base T=5~K). All SANS data presented in this manuscript correspond to differences between low-temperature measurements and high temperature backgrounds taken in the paramagnetic phase of each material. The resulting 2D datasets were analyzed using 1D cuts. Order parameters for the different scattering components were obtained by fitting the annular $\mathrm{Q}$-cuts to Lorentzian profiles. Likewise, I vs $\mathrm{Q}$ along specific fin directions were obtained using the rectangular cuts displayed in Supplementary Fig.~S-2, after the geometric correction detailed in Section~S-3 was applied to rocking curves. In MMO, these geometric corrections were complicated by the mosaicity inherent in the sample, and the intensity distributions represent cuts of 2D data taken with one orientation.

SQUID magnetometry was carried out inside a Quantum Design Magnetic Properties Measurement System (QD MPMS 3). DC measurements were performed for field and temperature sweeps using the standard DC scan mode. For time-dependent measurements, the vibrating sample magnetometer (VSM) option was used, in order to obtain the highest possible time resolution. To observe the time-dependent relaxations, field steps employed the highest available ramp rate (700 Oe/s).

Magnetocapacitance measurements were performed inside a Quantum Design Physical Properties Measurement System with a home-built insert at the National High Magnetic Field Laboratory at Los Alamos National Laboratory. To probe the magnetocapacitance effect, the capacitance of the MVO sample was measured with an Andeen-Hagerling AH-2500A commercial capacitance bridge at a driving frequency of 1000 Hz and amplitude of 15V. Silver paint and gold wire were used as contacts on the parallel faces of the plate-like (1.5mm $\times$ 3mm $\times$ 0.1 mm) MVO sample. A calibrated Cernox sensor near the sample was used to monitor temperature. The resistance was measured using a Keithley 6517A electrometer with a bias voltage of 10V, with the same mounting configuration and probe as the capacitance measurements. 

\section{Acknowledgements}
This work was performed under the support of the National Science Foundation, under grant number DMR-1455264-CAR. Research at the High Flux Isotope Reactor was sponsored by the U. S. Department of Energy, Office of Basic Energy Sciences, Scientific User Facilities Division.

The National High Magnetic Field Lab is funded by the U.S. National Science Foundation through Cooperative Grant No. DMR-1157490, the U.S. D.O.E and the State of Florida.

Small-angle scattering work was supported by the U.S. Department of Energy, Office of Science, Office of Workforce Development for Teachers and Scientists, Office of Science Graduate Student Research (SCGSR) program. The SCGSR program is administered by the Oak Ridge Institute for Science and Education (ORISE) for the DOE. ORISE is managed by ORAU under contract number DE-SC0014664. All opinions expressed in this paper are the author’s and do not necessarily reflect the policies and views of DOE, ORAU, or ORISE.

\section{Author Contributions}
A. T. and H. D. Z. grew the single crystal samples to be characterized. B. W., X. W., and R. B. performed the initial microscopy work. L. L. K., A. T., D. R. P., A. V. Z., A. A. A., L. D., and G. J. M. performed SANS experiments. L. L. K. performed (L. D. S. and K. C. L. provided insights into) the analysis and modelling of SANS data. L. L. K. performed magnetization studies. M. L. and V. S. Z. performed all magnetocapacitance and transport measurements. All authors contributed to interpreting results. The manuscript was written by L. L. K., A. A. A., L. D. S., and G. J. M. 

\section{Competing Interests}
The authors declare no competing interests.

\bibliographystyle{naturemag}
\bibliography{MyCollection_current}


\end{document}



\title{Supplementary Information: Domain wall patterning and giant response functions in ferrimagnetic spinels}

\author{L. L. Kish}
\email{lazark2@illinois.edu}
\affiliation{Department of Physics and Materials Research Laboratory, University of Illinois at Urbana-Champaign, Urbana, Illinois 61801, USA}

\author{A. Thaler}
\affiliation{Department of Physics and Materials Research Laboratory, University of Illinois at Urbana-Champaign, Urbana, Illinois 61801, USA}
\affiliation{Neutron Scattering Division, Oak Ridge National Laboratory, Oak Ridge, Tennessee 37831, USA}

\author{M. Lee}
\affiliation{National High Magnetic Field Laboratory, Los Alamos National Laboratory, Los Alamos, New Mexico 87544, USA}

\author{A. V. Zakrzewski}
\affiliation{Department of Physics and Materials Research Laboratory, University of Illinois at Urbana-Champaign, Urbana, Illinois 61801, USA}

\author{D. Reig-i-Plessis}
\affiliation{Department of Physics and Astronomy and Quantum Matter Institute, University of British Columbia, Vancouver, BC V6T 1Z1, Canada}

\author{B. Wolin}
\affiliation{Department of Physics and Materials Research Laboratory, University of Illinois at Urbana-Champaign, Urbana, Illinois 61801, USA}

\author{X. Wang}
\affiliation{Department of Physics and Materials Research Laboratory, University of Illinois at Urbana-Champaign, Urbana, Illinois 61801, USA}

\author{K.C. Littrell}
\affiliation{Neutron Scattering Division, Oak Ridge National Laboratory, Oak Ridge, Tennessee 37831, USA}

\author{R. Budakian}
\affiliation{Department of Physics and Materials Research Laboratory, University of Illinois at Urbana-Champaign, Urbana, Illinois 61801, USA}
\affiliation{Department of Physics and Astronomy, University of Waterloo, Waterloo, Ontario N2L 3G1, Canada}

\author{H. D. Zhou}
\affiliation{Department of Physics and Astronomy, University of Tennessee, Knoxville, Tennessee 37996, USA}

\author{V. S. Zapf}
\affiliation{National High Magnetic Field Laboratory, Los Alamos National Laboratory, Los Alamos, New Mexico 87544, USA}

\author{A. A. Aczel}
\affiliation{Neutron Scattering Division, Oak Ridge National Laboratory, Oak Ridge, Tennessee 37831, USA}

\author{L. DeBeer-Schmitt}
\affiliation{Neutron Scattering Division, Oak Ridge National Laboratory, Oak Ridge, Tennessee 37831, USA}

\author{G. J. MacDougall}
\affiliation{Department of Physics and Materials Research Laboratory, University of Illinois at Urbana-Champaign, Urbana, Illinois 61801, USA}

\date{\today}

\maketitle

\renewcommand{\thepage}{S-\arabic{page}}
\renewcommand{\thesection}{S-\arabic{section}}
\renewcommand{\thetable}{S-\arabic{table}}
\renewcommand{\thefigure}{S-\arabic{figure}}

\section{Overview of crystallographic conventions}\label{section:S_Dirs}

\begin{figure*}[htp]
	\centering
	\includegraphics[width=0.5\textwidth]{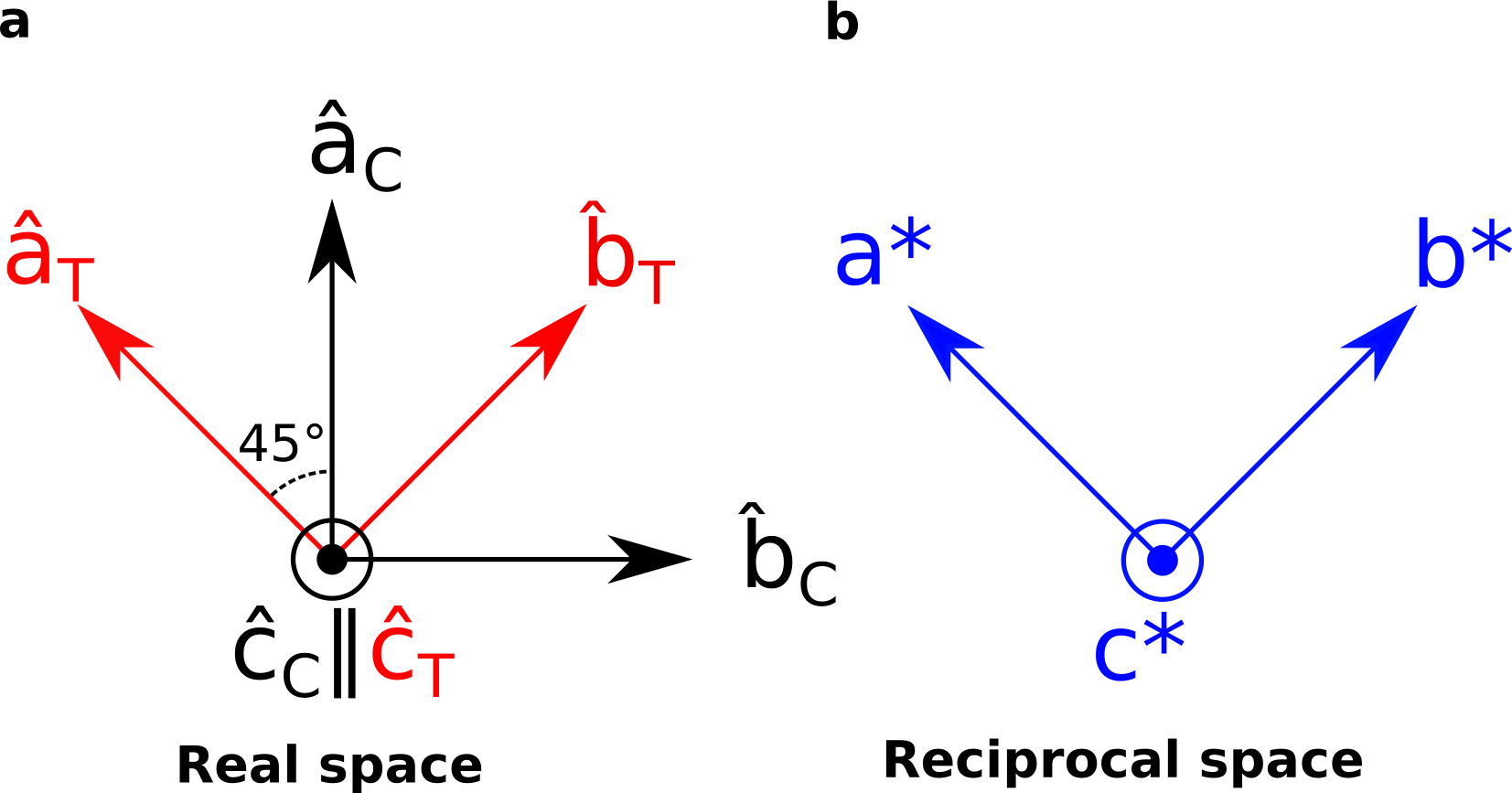}
	\caption{Schematic showing our conventions for labeling the different crystal directions. (\textbf{a}) Relationship between conventions for cubic and tetragonal unit cells. (\textbf{b}) Reciprocal space directions in the tetragonal convention.}\label{fig:S_Dirs}
\end{figure*}

Here we present an overview of the crystallographic conventions used in this paper. As stated in the text, the two materials do not share space groups at relevant temperatures. Mn$\mathrm{_3}$O$\mathrm{_4}$ (MMO) has the tetragonal spacegroup $\mathrm{I4_1/amd}$ at room temperature, and retains this symmetry until 33K whereupon it transitions to one of two orthorhombic phases. MnV$\mathrm{_2}$O$\mathrm{_4}$ (MVO) retains the cubic $\mathrm{Fd\bar{3}m}$ spacegroup until $\mathrm{53K}$, and then transitions to its low-temperature tetragonal $I4_1/a$ symmetry. We wish to highlight the differing conventions for the cubic and tetragonal unit cells. By definition, the tetragonal cell selects a unique axis $\mathrm{c_T}$ along one of three symmetry-equivalent cubic directions $\mathrm{c_C}$. The remaining tetragonal directions $\mathrm{a_T}$ and $\mathrm{b_T}$ are conventionally defined with a 45$^\circ$ rotation away from the cubic directions $\mathrm{a_C}$ and $\mathrm{b_C}$. This is demonstrated in Fig. \ref{fig:S_Dirs}. Because MMO starts out with a uniquely defined tetragonal $\mathrm{c_T}$ axis, the main text references all directions using the tetragonal convention. Although MVO in ambient conditions may be expected to retain a global cubic symmetry due to equal proportions of $\mathrm{c_T}$-axis variant domains, in-plane sample strain and magnetic field both serve to define a unique axis normal to the sample plane. We therefore choose our reference $\mathrm{c_T}$ to be parallel to the cubic $\mathrm{c_C}$ direction normal to the crystal plate's surface and our applied magnetic field. The remaining two directions $\mathrm{a_T}$ and $\mathrm{b_T}$ are defined using the tetragonal convention. For both materials, the reciprocal lattice vectors are parallel to the real-space directions, as shown in Fig. \ref{fig:S_Dirs}. For simplicity, the main text drops the ``T" subscript and respectively denotes the real and reciprocal space tetragonal directions as ``a,b,c" and ``$\mathrm{a^*,b^*,c^*}$".

\section{Rectangular line cuts for I vs Q curves}\label{section:S_RCuts}

\begin{figure*}[htp]
	\centering
	\includegraphics[width=.5\textwidth]{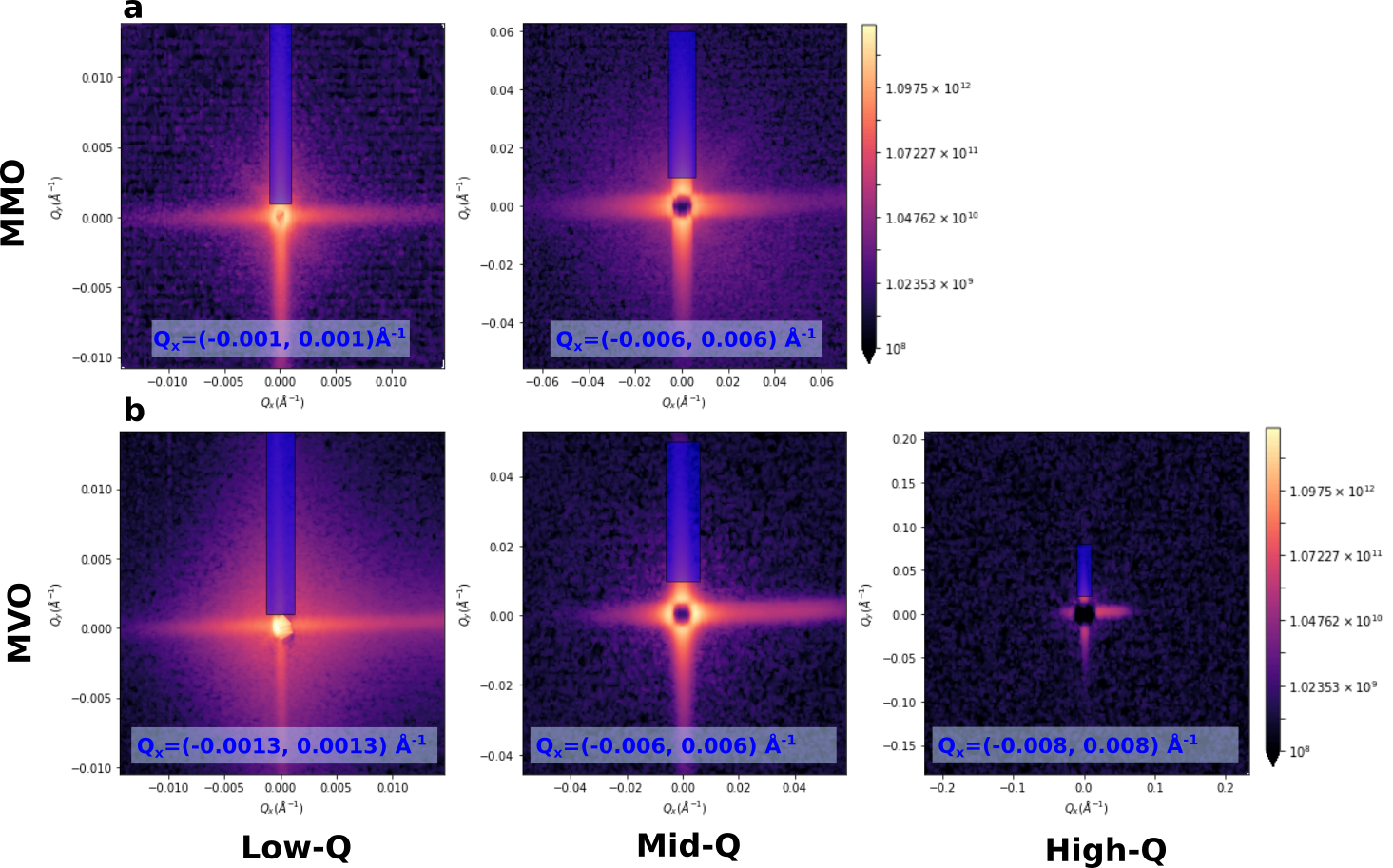}
	\caption{\textbf{Representations of SANS rectangular cuts.} Rectangular cuts for (\textbf{a}) MMO and (\textbf{b}) MVO, for each range of Q measured. The horizontal bounds of the integration regions are listed at the bottom of each panel.}\label{fig:S_RCuts}
\end{figure*}

To fit a subset of our 2D anisotropic datasets to 1D models, we took rectangular linecuts of the fin scattering parallel to the propagation direction of the feature. The regions of integration for these linecuts, with the transverse integration ranges for each instrument setting, are displayed by the blue rectangles in Fig. \ref{fig:S_RCuts}. The integration region was widened for instrument settings at progressively higher Q, in order to account for resolution broadening of the features.

\section{Geometric correction for anisotropic scattering}\label{section:S_Gcorr}
Single crystal samples have the potential to display sharply anisotropic scattering, where the geometric effect of Ewald sphere curvature is significant compared to the width of the feature. This must be accounted for in order to accurately fit these data without artificial drop-offs of the intensity with increasing Q. In the case of MMO, this issue is complicated by small misalignments in the grain structure, which result in diverging bands of intensity in the mid-Q and high-Q range.  

\begin{figure*}[htp]
	\centering
	\includegraphics[width=.5\textwidth]{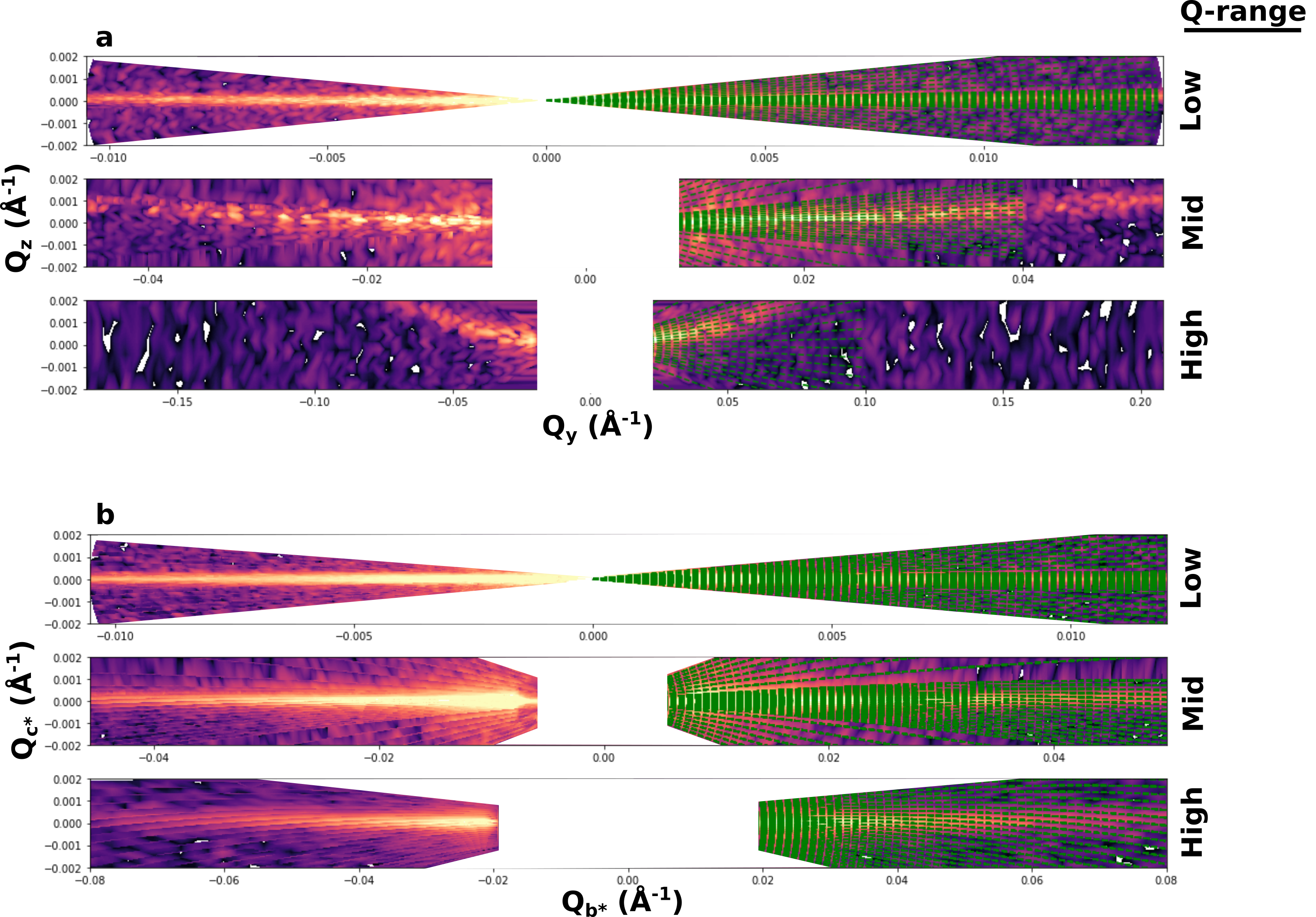}
	\caption{\textbf{Effects of Ewald sphere curvature and correction for MVO rectangular data.} The 2D panels represent intensity captured by the rectangular linecuts as a function of sample rotation in (\textbf{a}) detector coordinates and (\textbf{b}) projected onto reciprocal space crystal axes. The green dashed lines represent the detector curvature.}\label{fig:S_Curvature}
\end{figure*}

For MVO, where our crystals are truly single-grain, we are able to fully correct our data for this effect. This was achieved using finely-spaced scans as a function of sample rotation with respect to the incident beam (around the x-axis), to capture the 3-D nature of the scattering. The results are then projected onto the crystal axes in reciprocal space for line-cuts. The raw data for MVO, as collected by rectangular cuts in the yz-plane ($\mathrm{\hat{z}}$ normal to detector plane) and the resulting corrected projection are shown in \ref{fig:S_Curvature}, with the detector plane represented by the dashed green lines. 

\section{Modelling of SANS intensity vs Q}\label{section:S_Model}
Due to the relative magnitude of the magnetic moment in these samples compared with the expected atomic density fluctuations arising from structural distortions, we assume that nearly all of the SANS intensity is magnetic in origin.

For unpolarized SANS, the cross-section for magnetic scattering is proportional to the squared modulus of the Fourier-transform of the magnetization component perpendicular to $\mathrm{\hat{Q}}$: 
\begin{equation}
I_M(\vec{Q})\propto \left|\int d\textbf{r}\textbf{M}_{\perp\textbf{Q}}(\textbf{r})e^{i\textbf{Q}\cdot\textbf{r}}\right|^2
\end{equation}

These measurements cannot distinguish between two distinct microscopic pictures for the stripes. The first, which would correspond to order-disorder magnetic phase coexistence, involves a largely unidirectional magnetization with a spatially varying amplitude. The second picture, which corresponds more naturally to nanostructural twin domains, involves a spatially alternating magnetization direction from stripe to stripe as the magnetic easy axis rotates into the perpendicular direction for each variant of the c-axis structural distortion.

\begin{figure*}[htp]
	\centering
	\includegraphics[width=0.5\textwidth]{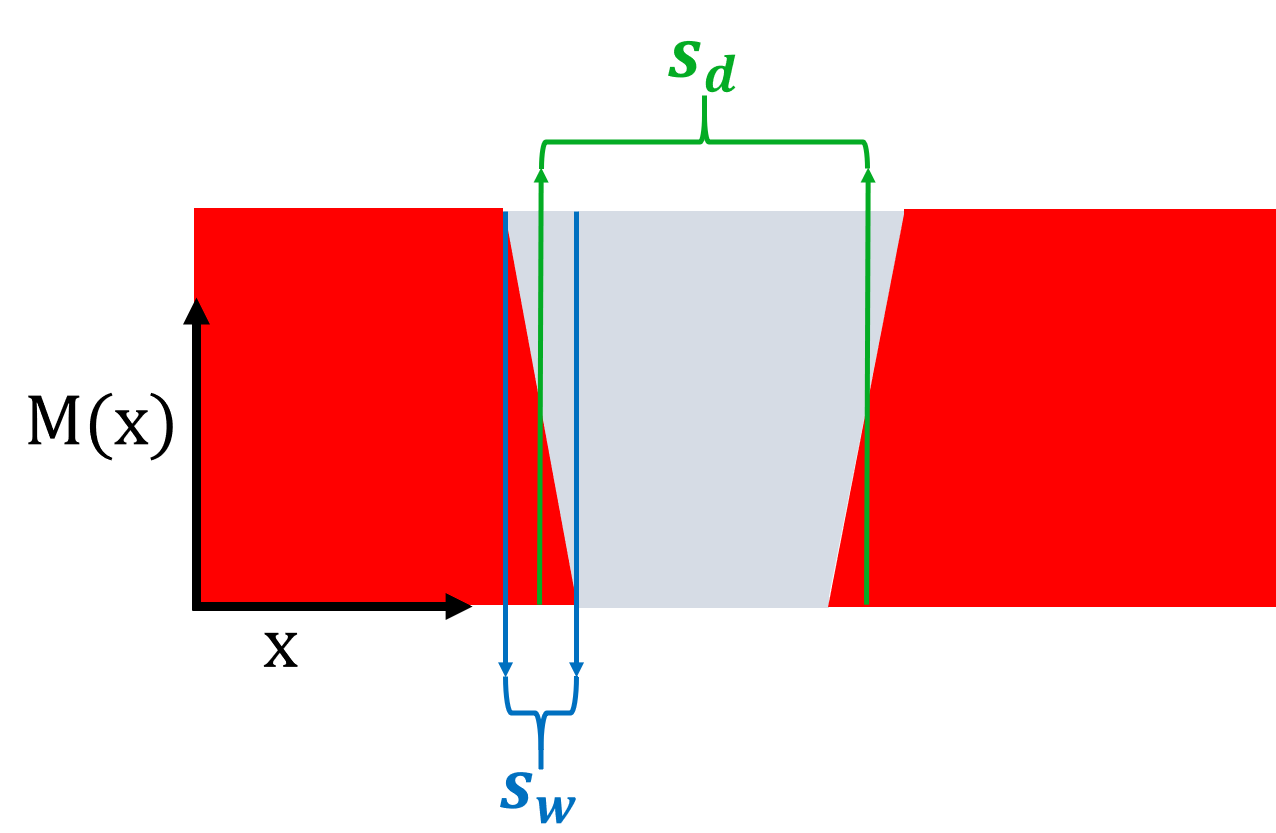}
	\caption{\textbf{Schematic representation of our model of a solitary stripe in real space.} $\mathrm{M(x)}$ represents a spatially-varying scalar component of the magnetization, with stripe width $\mathrm{s_d}$ and wall width $\mathrm{s_w}$.}\label{fig:S_Stripe_Cartoon}
\end{figure*}

Bearing in mind this ambiguity, we assume the former picture for fitting purposes, where the real-space variation of $\mathrm{|\textbf{M}|}$ is constant within the bulk of a domain, and changes linearly within the interfaces. This is represented schematically in Fig. \ref{fig:S_Stripe_Cartoon}, where the width of a stripe is $\mathrm{s_d}$ and the width of its walls are $\mathrm{s_w}$. By the convolution theorem, the Fourier-transform of the magnetization amplitude is simply the product of two sinc functions, and scattering from a stripe of this form can be written as:
\begin{equation}
\mathrm{I(Q_{b*})=a~sinc^2\left(\frac{s_d Q_{b*}}{2}\right)sinc^2\left(\frac{s_w Q_{b*}}{2}\right).}
\end{equation}
Here, $a$ is a proportionality constant. In order to take account variations in these lengthscales, each sinc$^2$ term is numerically convoluted with a Gaussian size distribution 
\begin{equation}
G=\frac{\exp(-\frac{(s-s_*)^2}{2\sigma^2})}{\sqrt{2\pi}\sigma},   
\end{equation} 
with a high-Q cut off at s=0:
\begin{equation}
\mathrm{I(Q_{b*})=a\left(sinc^2\left(\frac{s Q_{b*}}{2}\right) * G(s,s_d,\sigma_d)\right)\left(sinc^2\left(\frac{s Q_{b*}}{2}\right) * G(s,s_w,\sigma_w)\right).}
\label{eq:stripe_FF}
\end{equation}
For use in fitting intensities, this model is once more convoluted with the Gaussian resolution function of GP-SANS. In principle, inter-stripe correlations for densely packed stripe domains could generate peaks in the data. However, we found that the single stripe form factor described the variation in Q sufficiently well without additional structure factor components for most of our data. While the lack of a structure factor is a likely effect of high polydispersity, we cannot preclude the possibility of additional structure factor components below the low-Q limits of our measurement.
The exception to this was the case of MMO with $\mathrm{H||c^*}$, as displayed in the main text, as a weak peak appeared in the data with increasing field. Data in this case was modelled by a combination of Lorentzian and $\mathrm{Q^{-4}}$ intensity components, as our measured Q-range was insufficient to produce a unified fit involving structure and form factors. 

\section{Magnetic transitions and characterization}\label{section:S_MagChar}
\begin{figure*}[htp]
	\centering
	\includegraphics[width=.5\textwidth]{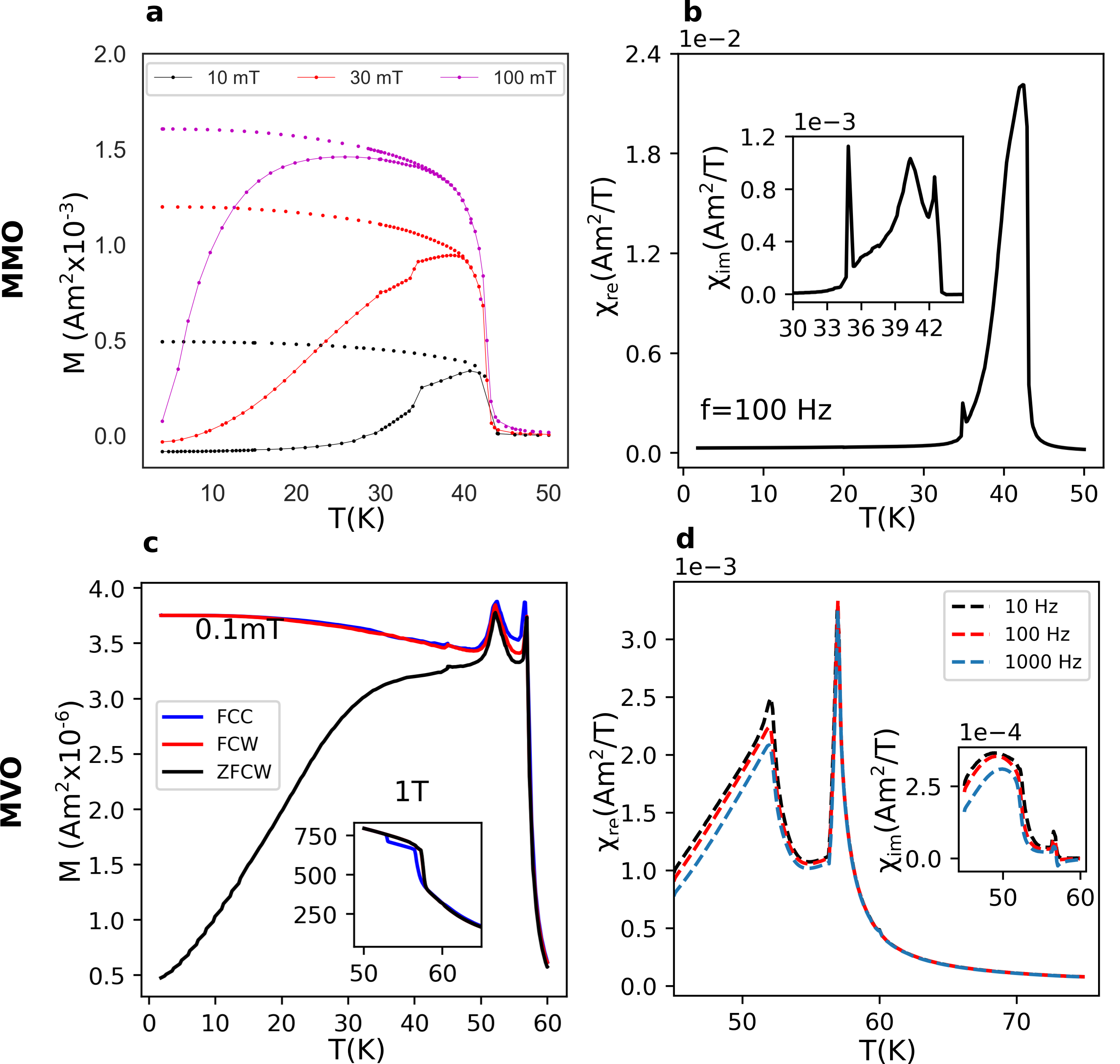}
	\caption{\textbf{Temperature dependent DC and AC magnetic measurements on MMO and MVO.} Plots show zero-field cooled warming (ZFCW), field-cooled cooling (FCC), and field-cooled warming (FCW) data. \textbf{a} MMO DC magnetization at various fields: ZFCW curves are connected lines, FCW are dotted. \textbf{b} MVO DC at low field (inset: high field) \textbf{c} MMO AC (ZFCW,$\mathrm{H_{AC}=0.01mT}$) susceptibility, \textbf{d} MVO AC (ZFCW, $\mathrm{H_{AC}=0.01mT}$) susceptibility at various frequencies.}\label{fig:S_MvT}
\end{figure*}

Temperature dependent DC (Fig.~\ref{fig:S_MvT} a,b) and AC (Fig.~\ref{fig:S_MvT} c,d) magnetization curves show the relevant magnetic transitions in each material. The magnetostructural transitions, which are associated with an increase of the local magnetic anisotropy, coincide as expected with an increased zero field-cooled-field-cooled splitting in the DC magnetization. The AC susceptibilities observe a peak at the transition then a subsequent drop. In MVO the AC response sees a degree of frequency dependence as well, implying a range of relaxation times associated with the orbital-ordering transition. This behavior is not replicated by the higher temperature spin-only ordering transition.

\section{Bulk magnetic response of MMO}\label{section:S_MagMMO}
\begin{figure*}[htp]
	\centering
	\includegraphics[width=.5\textwidth]{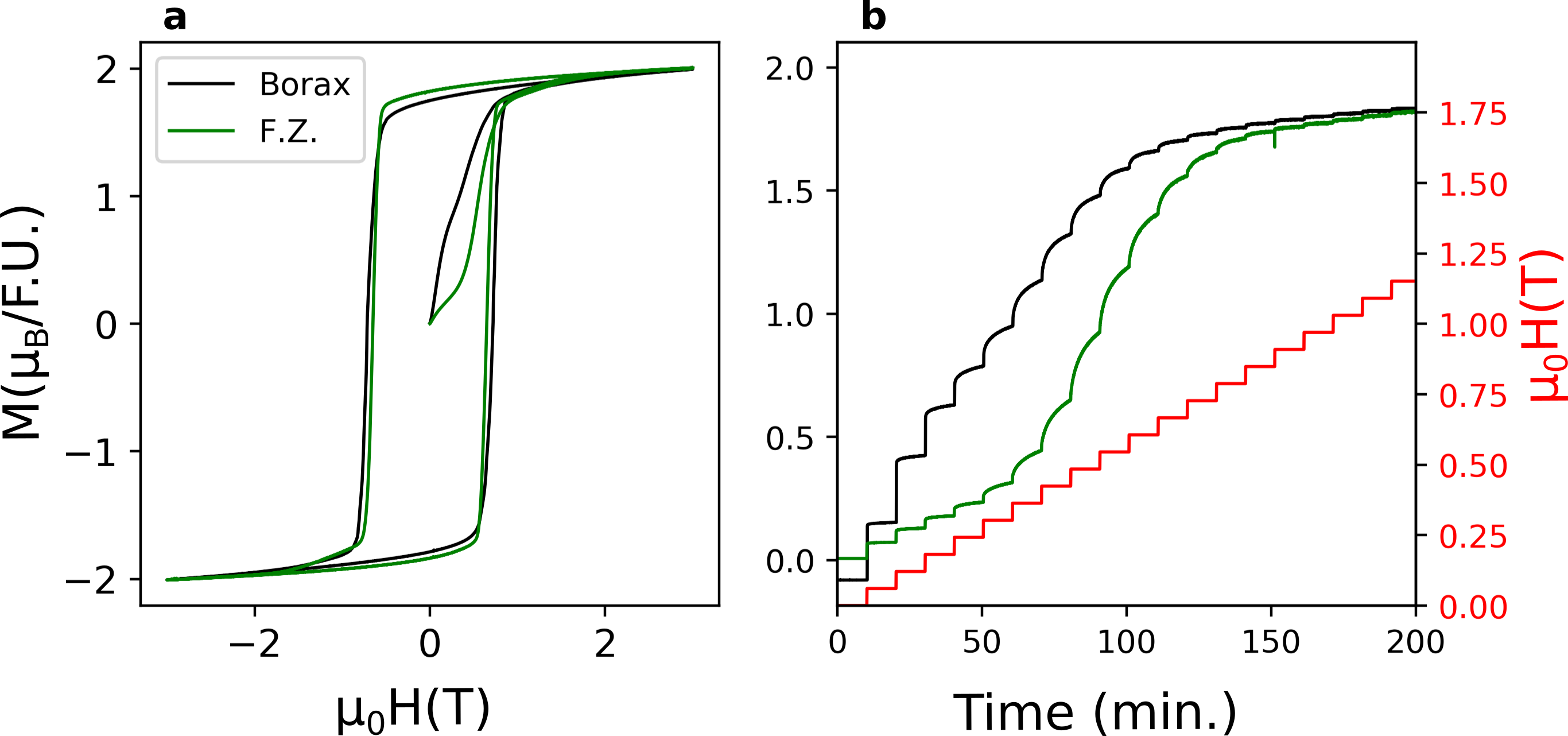}
	\caption{\textbf{MMO magnetization for two different growth methods.} Magnetization as a function of (\textbf{a}) field and (\textbf{b}) time for MMO, comparing samples grown using floating-zone and Borax flux methods.}\label{fig:S_MMO_M}
\end{figure*}

In the main text, we associate time-dependent behavior in the virgin hysteresis loops of MVO with motion of magnetostructural domain walls. The prevalence of this region in the hysteresis loop was associated with sample strain. In Fig. \ref{fig:S_MMO_M} we present analogous magnetization data for MMO, for both high strain floating-zone (F.Z.) and low strain Borax growths. 

For the sake of comparison, measured Borax moments were rescaled to match the F.Z. extrema. The hysteresis loops are quite similar, with the largest difference once again occurring in the virgin curve. As observed for MVO, a kink in the virgin curve is associated with a crossover in the time dependent magnetization from a fast to a slowly relaxing timescale. It is worth noting that the Borax sample was attached via epoxy to a sample holder for this measurement due to sample geometry. This likely induced mounting stress comparable to the glued case in our MVO measurements. 

\section{SANS from low-strain MMO sample}\label{section:S_SANS_Flux}
\begin{figure*}[htp]
	\centering
	\includegraphics[width=.5\textwidth]{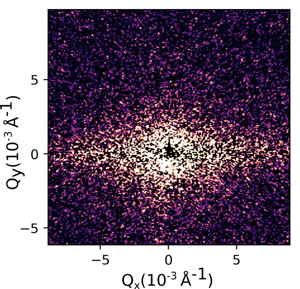}
	\caption{\textbf{SANS from Borax sample growth.} Measurement was done at T=1.5K. The plot shows the difference with a high temperature dataset taken in the paramagnetic regime of MMO at T=49K.}\label{fig:S_Flux}
\end{figure*}

Our SANS characterization of Borax samples failed to locate any sign of the temperature-dependent fin scattering. Fig. \ref{fig:S_Flux} displays a long-count difference pattern between base temperature (T=1.5K) and background at (T=49K). The horizontal anisotropic scattering is due to small-angle reflections from the edges of our crystal, and shows none of the characteristic temperature dependence of the stripe domains. Although we cannot say that stripe domains are absent in this crystal, at the very least we demonstrate a dramatic suppression of the effect. 

\section{Complementary capapacitance data for MVO}\label{section:S_Cap}
\begin{figure*}[htp]
	\centering
	\includegraphics[width=.5\textwidth]{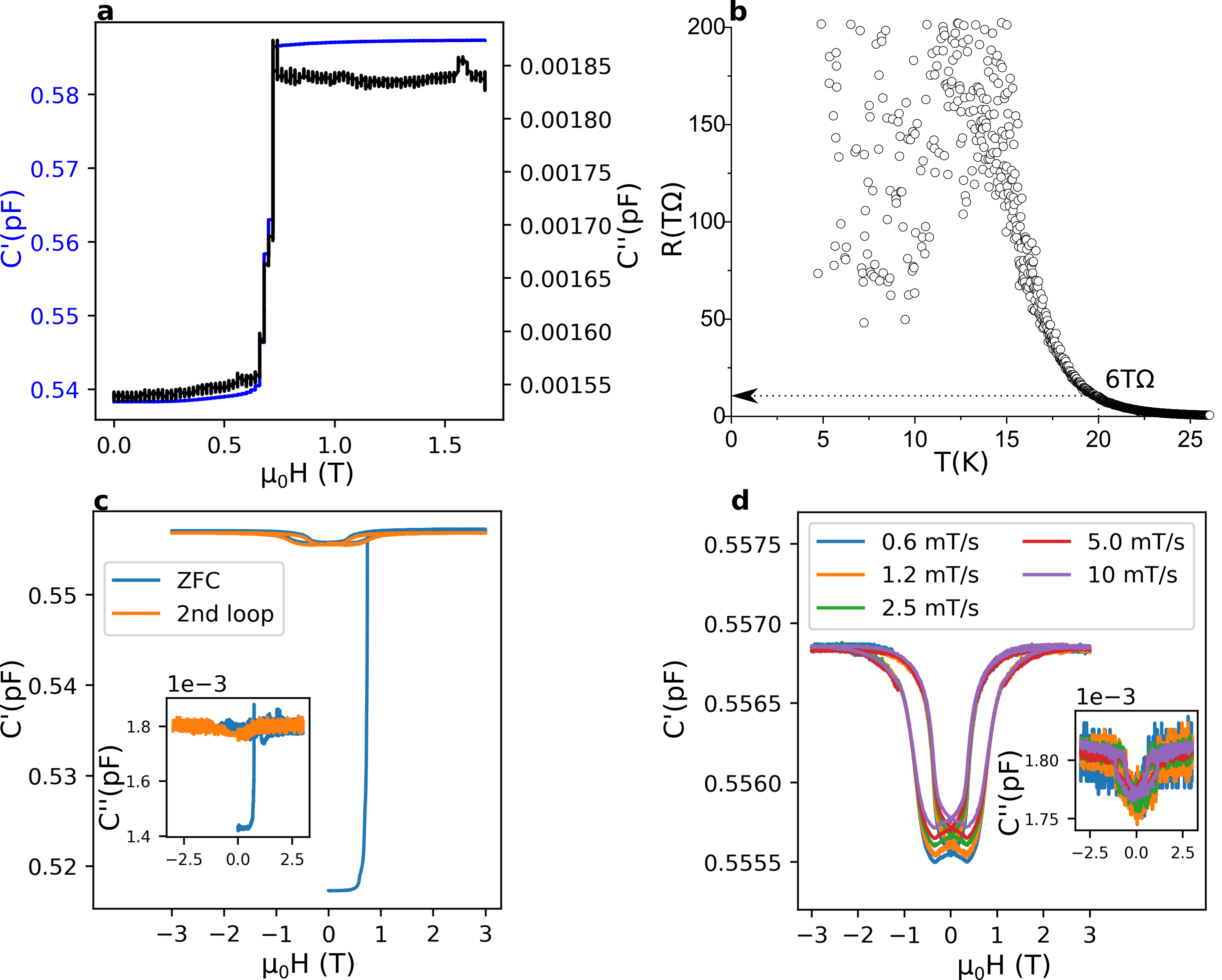}
	\caption{\textbf{Complementary capacitance data as a function of field and time for MVO.} Measurements were done at T=3K. C' refers to the in-phase signal, while C'' refers to the out-of-phase signal. (\textbf{a}) Raw data for the time-dependent measurement in the main text. C'' shows a jump of similar to that in C'. (\textbf{b}) Temperature dependence of sample resistance. (\textbf{c}) Comparison of virgin curve and further hysteresis loops. (\textbf{d}) Ramping rate dependence of hysteresis loops, showing systematic reduction of the size of the loop with faster rates.}\label{fig:S_MVO_C}
\end{figure*}

In this section we present complementary capacitance measurements on MVO at T=3K. \ref{fig:S_MVO_C}a displays the raw in (out of) phase capacitance signals C'(C'') as a function of field, which led to the results shown in the main text. C'' displays a similar jump to C', though both curves likely incorporate the effects of changes in sample geometry.  \ref{fig:S_MVO_C}b shows sample resistance as a function of temperature as referred to from the main text, exceeding 6T$\Omega$ at 20K, and increasing beyond our measurement limit at lower temperatures. The low-temperature scatter is insulating background noise. Fig. \ref{fig:S_MVO_C}c compares the virgin curve and subsequent capacitive hysteresis loops. The maximum change in the field-cycled loops is a factor of 60 less than that of the virgin curve. Hysteresis loops at various field ramp rates are displayed in \ref{fig:S_MVO_C}d, showing a small systematic reduction in the size of the response with increased rate. This indicates that MVO retains a small degree of time-dependent behavior in its field-cycled state. 

